# Laser-driven electrodynamic implosion of fast ions in a thin shell

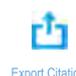

View Online    Export Citation    CrossMark

S. Yu. Gus'kov,[1] Ph. Korneev,[1,a] and M. Murakami[2]

**AFFILIATIONS**

[1] P. N. Lebedev Physical Institute of Russian Academy of Sciences, 53 Leninskii Prospect, 119991 Moscow, Russian Federation
[2] Institute of Laser Engineering, Osaka University, 565-0871 Osaka, Japan

[a] Author to whom correspondence should be addressed: korneev@theor.mephi.ru

**ABSTRACT**

Collision of laser-driven subrelativistic high-density ion flows provides a way to create extremely compressed ion conglomerates and study their properties. This paper presents a theoretical study of the electrodynamic implosion of ions inside a hollow spherical or cylindrical shell irradiated by femtosecond petawatt laser pulses. We propose to apply a very effective mechanism for ion acceleration in a self-consistent field with strong charge separation, based on the oscillation of laser-accelerated fast electrons in this field near the thin shell. Fast electrons are generated on the outer side of the shell under irradiation by the intense laser pulses. It is shown that ions, in particular protons, may be accelerated at the implosion stage to energies of tens and hundreds of MeV when a sub-micrometer shell is irradiated by femtosecond laser pulses with an intensity of $10^{21}$–$10^{23}$ W cm$^{-2}$.

## I. INTRODUCTION

Generation of laser-accelerated charged particles is one of the most interesting topics in high-energy-density physics. In addition to the fundamental significance of the phenomenon, there are many important applications related to the creation of compact powerful sources of charged particle beams as diagnostic tools, initiators of ignition in inertial confinement fusion (ICF), injectors for accelerators, and components of equipment for technological and medical applications. With regard to aspects of fundamental physics in relation to laser-produced fast-ion beams, it is necessary to note their great potential as effective tools for implementing fast ignition in different ICF schemes[1–3] and stimulating various types of nuclear reactions,[4–6] as well as for diagnostics of dense plasma objects and laboratory astrophysics. The electrodynamics of the generation, acceleration, and interaction of laser-produced fast ions are particularly important when special targets such as hollow spherical or cylindrical shells, hemispherical shells, or conical hollow targets are used to provide focusing and collision of high-energy ion beams. Collisions of high-density and high-energy ion flows inside a shell target give rise to a variety of interesting and important high-energy-density physical phenomena, thereby providing a useful way to study such phenomena in detail. It should additionally be noted that the use of hemispherical, cylindrical, or conical shell targets enables the investigation of electrodynamic implosions of fast ions in an open geometry convenient for diagnostics.

In ICF studies, there is a long history of experiments on generation of nonstationary plasmas via collisions between high-density, high-energy ion flows inside shell targets. There are two well-known schemes for such experiments with thin shells irradiated by laser pulses. Both of these were developed in the 1980s with laser intensities of $10^{15}$–$10^{17}$ W cm$^{-2}$. One was widely used in experiments on the Nova installation (LLNL, USA) and was called the "exploding pusher" (see the review in Ref. 7 and the references therein). It involved heating the outer surface of a thin shell, for example, glass or polystyrene with a thickness of about 1 $\mu$m or less, with a pulse from an Nd-laser with a duration of about 1 ns. The shell could be hollow or filled with a gas mixture containing hydrogen isotopes, depending on the purposes of the particular experiment. Owing to the rapid heating of the shell over its entire thickness by a wave of electronic thermal conductivity and fast electrons, a thermal explosion occurred. At this stage, about half of the shell was flying out, while the inner part imploded inside the shell's cavity, forming a converging supersonic plasma stream toward the center.

The other approach was used in experiments on the Iskra-5 facility (RFNC-VNIIEF, Russia) and was called the "inverse corona" (see the review in Ref. 8 and the references therein). The target used was a hollow shell with special entrance holes for the laser beams. The inner surfaces of the shell was covered with a layer of material, for example, deuterated polyethylene ($CD_2$). Beams from iodine lasers were focused on the inner surface of the shell. Rapid heating of the material on the inner surface of the shell (the duration of the laser pulse was about 0.4 ns) led to the formation of a supersonic flow of laser-produced plasma (as in the "exploding pusher" scheme), converging toward the center of the target. Experiments using "exploding pusher" and "inverse corona" schemes provided useful data for plasma physics, including data on the isentropes of various substances, on the cumulative supersonic flows and their interaction, on the relaxation processes leading to the formation of a nonstationary plasma, and on the properties of high-temperature (>10 keV) plasmas.

In recent years, the availability of short petawatt pulses has made possible studies of the interaction of electrodynamically accelerated ion flows with velocities about two orders of magnitude higher than those of thermal origin. The interaction of such subrelativistic ion flows in the absence of a counter-pressure can lead to the formation of an extremely compressed ion conglomerate. The exploitation of this phenomenon to generate accelerated ions inside a microscale cavity in a quasistationary electrostatic field was theoretically investigated in Refs. 9 and 10. It was predicted that it should be possible to accelerate protons to an energy of several hundred MeV and compress ions to a density significantly exceeding that of a solid.

This paper is devoted to a theoretical study of the electrodynamic acceleration of ions inside a hollow spherical or cylindrical shell irradiated by intense short laser pulses. This situation differs from that in earlier studies in its somewhat lower symmetry, owing to the finite number of laser beams. However, this approach may provide a significantly greater number of imploding ions, which is important for further studies of dense and hot ion ensembles. In its first stage, the proposed approach utilizes a very effective mechanism for ion acceleration in a self-consistent field of strongly separated charges. This field is formed by laser-accelerated fast electrons, captured near the thin shell and oscillating there.[11] Fast electrons are generated on the outer side of the shell when it is irradiated by a laser pulse with a relativistic intensity. Most of them are then captured near the shell and create a strong potential; see the scheme in Fig. 1. It is shown below that under the action of femtosecond laser pulses with intensities of $10^{21}$–$10^{23}$ W cm$^{-2}$, ions, in particular protons, can be accelerated inside the shell to energies of tens and hundreds of MeV. Section II presents a model describing the characteristics of a separated charge field caused by captured laser-produced fast electrons oscillating near the thin wall of a spherical shell. In Sec. III, the dependences of the energies of ions electrodynamically accelerated inside the hollow shell on the parameters of the shell and of the femtosecond petawatt laser pulse are estimated and discussed. In Sec. IV, the laser-driven electrodynamic implosion is demonstrated on the basis of the results of two-dimensional particle-in-cell (2D PIC) simulations.

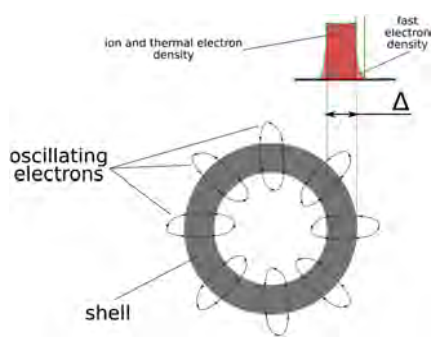

**FIG. 1.** Schematic illustration of the formation of a strong self-consistent potential, accelerating ions to multi-MeV energies toward the center of the shell, heated by symmetrical irradiation with short intense laser pulses. The thicknesses of the shell and Debye layer are denoted by $\Delta$ and $L_D$, respectively.

## II. SEPARATED CHARGE FIELD FORMED BY LASER-PRODUCED FAST ELECTRONS CAPTURED NEAR THE THIN WALL OF A SPHERICAL SHELL

The acceleration of ions in the field of separated charge formed by the oscillation of laser-accelerated fast electrons near a thin target is significantly more efficient than their acceleration by the traditional target normal sheath acceleration (TNSA) mechanism[4,12–17] in a thick target. This increased efficiency is due to the amplification of the accelerating field by multiple oscillations of fast electrons near a thin target[11] in comparison with a thick target, where a fast electron makes only one pass in the Debye layer, after which it returns to the solid target and thermalizes there. At a constant rate of production of fast electrons, the multiple oscillations of these electrons near the thin target result in an increase in their density near the shell wall, which in turn leads to a decrease in the thickness of the Debye layer. It is obvious that the target thickness $\Delta$ must meet two requirements to realize the effective electrodynamic acceleration of ions in the field of the oscillating trapped fast electrons. First, the energy loss of the fast electrons in the solid target must be negligible during the multiple oscillations. This means that the target thickness increased by a factor of the number $n$ of fast-electron oscillations must be significantly less than the mean free pass of a fast electron: $n\Delta \ll \lambda_c$. Second, to ensure reliable capture of fast electrons in the separated charge field, the target must not be too thin. As the Debye layer thickness (hereinafter called the Debye length) decreases during the oscillations of the captured fast electrons, it is sufficient that the target thickness be greater than the thick-shell Debye length, which occurs at the beginning of the process: $\Delta \geq L_{D0}$.

The maximum number of fast-electron oscillations is limited by the lifetime of the Debye layer, which is determined by the dynamics of relative motion of the fast electrons and the accelerated ions. Indeed, the full cycle of ion acceleration in this simplified description can be roughly divided into two stages. First, surface ions accelerate to an energy $\varepsilon_i \sim Z\varepsilon_h$, after they have passed a distance equal to the Debye length $L_D$ during the characteristic time $t_a \sim L_D/(Z\varepsilon_h/m_i)^{1/2}$ [here, $\varepsilon_h$ is the fast-electron energy, $L_D = (\varepsilon_h/4\pi n_D e^2)^{1/2}$, $n_D$ is the characteristic density of the fast electrons in the Debye layer, $e$ is the absolute value of the electron charge, $Ze$ is the ion charge, and $m_e$ and $m_i$ are the electron and ion masses, respectively]. At this stage, the Debye layer can be considered to be static, since it is formed in front of the accelerated ion flow, which moves much more slowly than the fast electrons. Then, a small proportion of the most energetic ions acquire energies close to the maximum possible value $\varepsilon_{i(m)} = (m_i/m_e)\varepsilon_h$. At this stage, the formation of the Debye layer has a more complex, dynamic character, since this layer is formed in front of the ion flow, the velocity of which becomes comparable to that of the fast electrons. Roughly, the acceleration ends with termination of the charge separation process, when the fast electrons and accelerated ions form a quasi-neutral flow. The characteristic time of this process is a factor of $(m_i/Zm_e)^{1/2}$ larger than the time $t_a$.

In Ref. 11, it was shown that in a target of light, hydrogen-containing materials, the condition of small energy losses of fast electrons in a solid target for the full ion acceleration cycle is fulfilled for relativistic fast electrons formed when the target is irradiated with a laser pulse with an intensity of $I_L \geq 10^{19}$ W cm$^{-2}$. For a laser pulse with the specified intensity, the range of target thicknesses that provide reliable capture of fast electrons and their insignificant energy losses in a solid target is significantly wider: 1–15 $\mu$m.

Consider the electrostatic field created by charge separation between fast electrons oscillating near the thin wall of a spherical shell of radius $R$ that significantly exceeds the thickness of the Debye layer, $R \gg L_D$. Hereinafter we assume that the Debye length is greater than the skin depth, which is usually true for relativistic electron plasmas.[18] The field strength is determined by the Debye length, which, for a given energy of fast electrons, depends only on their density. Following Ref. 11, for an estimate, assume that fast electrons are monoenergetic with energy and velocity $\varepsilon_h$ and $V_h$, respectively. The laser pulse intensity $I_L$ determines the energy density flux of fast electrons $I_h$ through the conversion efficiency $\eta$ of the laser energy transformation to fast-electron energy as $I_h = \eta I_L$. Let the fast electrons born at time $t$ be uniformly distributed over a volume $\Omega = 4\pi R^2(\Delta + 2L_D)$ in the region of their motion. Then, according to the solution presented in Ref. 11, at times significantly exceeding the duration of fast-electron oscillation, the fast-electron density in the Debye layer of a thin shell increases according to

$$n_D(t) = n_{D0} \frac{L_{D0}}{\Delta} \frac{t}{t_{D0}}. \quad (1)$$

In (1), $n_{D0}$ is the density of fast electrons in the Debye layer of a reference thick target of the same material, i.e., one in which the fast electrons return to the target and thermalize there after just a single turn in the Debye layer:

$$n_{D0} = \frac{2\eta I_L}{\varepsilon_h V_h}. \quad (2)$$

$t_{D0}$ is the time of flight of fast electrons in the Debye layer (hereinafter called the Debye time) of a thick target:

$$t_{D0} \equiv \frac{2L_{D0}}{V_h}. \quad (3)$$

$L_{D0}$ is the thick-shell Debye length at the very beginning of the interaction process, which is equal to the Debye length of a thick target:

$$L_{D0} = \left(\frac{\varepsilon_h}{4\pi n_{D0} e^2}\right)^{1/2}. \quad (4)$$

In turn, the Debye length $L_D$ decreases, and the strength of the separated charge field $E$ increases during fast-electron oscillations as follows:[11]

$$L_D(t) = L_{D0}\left(\frac{\Delta}{L_{D0}}\right)^{1/2}\left(\frac{t_{D0}}{t}\right)^{1/2},$$
$$E(t) = E_0\left(\frac{L_{D0}}{\Delta}\right)^{1/2}\left(\frac{t}{t_{D0}}\right)^{1/2}, \quad (5)$$

where $E_0$ is the field strength in the Debye layer of a thick target,

$$E_0 = \frac{\varepsilon_h}{eL_{D0}}. \quad (6)$$

Below, we consider relativistic fast electrons, for which, as shown above, there are wider possibilities for the field amplification effect in a thin target. For this case, the Debye length $L_{D0}$ and Debye time $t_{D0}$ were determined in Ref. 11 using the well-known scaling for the characteristic energy of laser-accelerated relativistic fast electrons:[19]

$$\varepsilon_h \approx 1.2\left(I_{L(19)}\lambda_\mu^2\right)^{1/2} (\text{MeV}), \qquad I_{L(19)}\lambda_\mu^2 > 1, \quad (7)$$

where $I_{L(19)}$ is the laser intensity measured in $10^{19}$ W cm$^{-2}$ and $\lambda_\mu$ is the laser radiation wavelength measured in micrometers. The expressions for these values are

$$L_{D0} \equiv \left(\frac{c\varepsilon_h^2}{4\pi\eta I_L e^2}\right)^{1/2} \approx 1.4 \times 10^{-5} \left(\frac{\lambda_\mu}{\eta}\right)^{1/2} (\text{cm}), \quad (8)$$

$$t_{D0} \equiv \left(\frac{\varepsilon_h^2}{2\pi\eta I_L e^2 c}\right)^{1/2} \approx 10^{-15}\left(\frac{\lambda_\mu}{\eta}\right)^{1/2} (\text{s}). \quad (9)$$

These do not depend on laser intensity and grow with increasing laser radiation wavelength as $\lambda^{1/2}$. For the first harmonic of the Nd-laser radiation at the laser energy conversion $\eta = 0.2$, the Debye length is about 0.3 $\mu$m and the Debye time is about 2 fs.

We now estimate the scales of fast-ion-flow characteristics related to electrodynamic implosion of a thin shell for the initial stage of ion acceleration, i.e., during the time $t_a = 2L_{D0}/V_i$ (where $V_i$ is the fast-ion velocity), when the Debye layer is almost static. For a thin shell with a wall thickness of $\Delta \gtrsim L_{D0}$, the formation of the static Debye layer can be considered in a one-dimensional geometry, when the front width scale is less than the Debye length. Owing to the effect of the field amplification with a decrease in the Debye length of fast electrons oscillating near the wall of the thin shell, fast ions over the time $t_a$ will acquire an energy exceeding the value of $Z\varepsilon_h$ by a ratio of $L_{D0}/L_D(t_a)$. Using (5) to determine this ratio $L_{D0}/L_D(t_a)$, for the scale of the energy of fast ions imploded inside the shell in the considered case of relativistic fast electrons, we obtain

$$\varepsilon_i = \left(\frac{L_{D0}}{\Delta}\right)^{1/2}\left(\frac{c}{V_i}\right)^{1/2} Z\varepsilon_h. \quad (10)$$

Then, for the velocity and energy of accelerated ion in the case of a laser pulse with relativistic intensity ($I_L > 10^{19}$ W cm$^{-2}$), the final result is

14 August 2023 07:37:06

$$V_i = 2^{2/5} c \left(\frac{L_{D0}}{\Delta}\right)^{1/5} \left(\frac{Z\varepsilon_h}{m_i c^2}\right)^{2/5}, \quad V_i \lesssim c, \tag{11}$$

$$\varepsilon_i = \frac{1}{2^{1/5}} \left(\frac{L_{D0}}{\Delta}\right)^{2/5} (m_i c^2)^{1/5} (Z\varepsilon_h)^{4/5}. \tag{12}$$

The number of ions accelerated inside the shell can be estimated from the charge-neutrality condition in the acceleration region based on the fact that their charge is equal to the charge of fast electrons located in the Debye layer at the time $t_a$, that is, $N_i = 4\pi R^2 n_D L_D / Z$. Using (1) and (11) to determine $n_D(t_a)$ and $L_D(t_a)$, this number may be rewritten as

$$N_i = \frac{4\pi R^2 n_{D0} L_{D0}}{2^{1/5} Z} \left(\frac{L_{D0}}{\Delta}\right)^{2/5} \left(\frac{m_i c^2}{Z\varepsilon_h}\right)^{1/5}, \tag{13}$$

where $n_{D0}$ and $L_{D0}$ are given by (2) and (4), respectively.

In Sec. III, the target and laser pulse parameters necessary for effective electrodynamic implosion of fast ions inside a thin shell are determined.

## III. LASER PULSE AND SHELL PARAMETERS FOR FAST-ION IMPLOSION IN THE FIELD OF OSCILLATING LASER-ACCELERATED FAST ELECTRONS

Consider a shell with an extremely small wall thickness equal to the thick-shell Debye length $L_{D0}$, which at the same time means that it is thick enough to retain electrons and stay within the framework of the considered model. Such a shell provides a reliable capture of fast electrons with negligible losses of their energy in the shell wall, maximum amplification of the separated charge field due to fast-electron oscillations, and a high resultant energy of the imploding ions. The characteristic values of the energy $\varepsilon_i$ and the number of accelerated ions $N_s$ are defined from (12) and (13) with the use of (2) and (4) for $n_{D0}$ and $L_{D0}$ and the scaling (7) for the fast-electron energy $\varepsilon_h$:

$$\varepsilon_i = 3.9 Z^{4/5} (I_{L(19)} \lambda_\mu^2)^{2/5} \text{ (MeV)}, \tag{14}$$

$$N_i = 1.8 \times 10^{18} \frac{R^2 I_{L(19)}^{2/5}}{Z^{6/5}} \frac{\eta^{1/2}}{\lambda_\mu^{7/10}}. \tag{15}$$

The duration of the laser pulse that generates fast electrons during the ion acceleration time $t_a$, according to (8) and (11), is

$$\tau = 1.8 \times 10^{-14} \frac{\lambda_\mu^{1/10}}{Z^{2/5} I_{L(19)}^{1/5} \eta^{1/2}} \text{ (s)}. \tag{16}$$

Let the shell radius be adjusted to the laser focal spot. Then,

$$R \equiv \left(\frac{E_L}{4\pi I_L \tau}\right)^{1/2} = 9 \times 10^{-4} Z^{1/5} \frac{E_L^{1/2} \eta^{1/4}}{I_{L(19)}^{2/5} \lambda_\mu^{1/20}} \text{ (cm)}. \tag{17}$$

Using (17), the number and flux density of accelerated ions are

$$N_i = 1.5 \times 10^{12} \frac{E_L \eta}{Z^{4/5} I_{L(19)}^{2/5} \lambda^{4/5}}, \tag{18}$$

$$W_i = 1.4 \times 10^{26} \frac{E_L \eta^{3/2}}{Z^{2/5} I_{L(19)}^{1/5} \lambda_\mu^{9/10}} \text{ (ions s}^{-1}\text{)}. \tag{19}$$

The expressions (14), (16), and (17) show that effective electrodynamic implosion of fast ions with energies from several MeV to several tens of MeV occurs when a hollow shell with a radius of several micrometers and a wall thickness of several tenths of a micrometer is irradiated by a femtosecond laser pulse with an intensity of $10^{19}$–$10^{23}$ W cm$^{-2}$. The shell radius and the optimal duration of the laser pulse decrease quite slightly with increasing pulse intensity as $I_L^{-2/5}$ and $I_L^{-1/5}$, respectively. These values are practically independent of laser wavelength, although the energy of the accelerated ions increases with wavelength as $\lambda^{4/5}$. The shell radius increases with growth of the laser energy as $E_L^{1/2}$, while the optimal pulse duration, which acts only at the stage of effective acceleration of ions, does not depend on the laser pulse energy. The fast-ion flow density under the specified conditions is ~ $10^{24}$ to $10^{25}$ particles per second. It grows linearly with the energy of the laser pulse and at the given laser to fast-electron energy conversion is almost inversely proportional to the laser wavelength. Later in the implosion process, the acquired kinetic energy of the imploding ions is transformed into their thermal motion after the central collision starts. Owing to thermalization, this energy is redistributed, and the resulting ion average energy in the central region at the time of their maximum density, which may be called the "effective temperature," is one or two orders of magnitude less, which is of the sub-MeV level. The redistribution is visualized with numerical simulations in Fig. 3.

**TABLE I.** Parameters of matched laser pulse and shell in the range of laser intensity of the first harmonic of the Nd-laser radiation $10^{19}$–$10^{23}$ W cm$^{-2}$, together with characteristic values of the energy, number, and generation rate of fast ions electrodynamically imploded inside the shell at the laser energy conversion to fast-electron energy $\eta = 0.2$. Here $I_L$ is the intensity, $E_L$ is the energy, $\tau_L$ is the duration of one of the laser pulses, $R$ is the radius and $\Delta$ is the thickness of the shell, $\varepsilon_i$ is the estimated energy, $N_i$ is the estimated number and $W_i$ is the estimated flux of the imploding ions. Boldface is the main case considered and illustrated in Figs. 2 and 3; boldface and italic are also the cases illustrated in Fig. 4.

| $I_L$ ($10^{19}$ W cm$^{-2}$) | $E_L$ (J) | $\tau_L$ (fs) | $R$ ($\mu$m) | $\Delta$ ($\mu$m) | $\varepsilon_i$ (MeV) | $N_i$ ($10^{10}$) | $W_i$ ($10^{24}$ ions s$^{-1}$) |
|---|---|---|---|---|---|---|---|
| 1 | 1 | 21.5 | 6.1 | 0.3 | 3.9 | 31 | 14 |
| *1* | *10* | *21.5* | *19.2* | *0.3* | *3.9* | *3.1 × 10$^2$* | *1.4 × 10$^2$* |
| 10 | 1 | 14.3 | 2.5 | 0.3 | 10 | 12 | 8.6 |
| **10** | **10** | **14.3** | **7.7** | **0.3** | **10** | **1.2 × 10$^2$** | **86** |
| 10 | 10$^2$ | 14.3 | 24.6 | 0.3 | 10 | 1.2 × 10$^3$ | 8.6 × 10$^2$ |
| *10$^2$* | *1* | *9.1* | *1* | *0.3* | *25* | *5* | *5.2* |
| 10$^2$ | 10 | 9.1 | 3.1 | 0.3 | 25 | 50 | 52 |
| 10$^2$ | 10$^2$ | 9.1 | 10 | 0.3 | 25 | 5 × 10$^2$ | 5.2 × 10$^2$ |
| 10$^3$ | 10 | 5.6 | 1.3 | 0.3 | 62 | 19 | 35 |
| 10$^3$ | 10$^2$ | 5.6 | 3.8 | 0.3 | 62 | 1.9 × 10$^2$ | 3.5 × 10$^2$ |
| 10$^4$ | 10 | 3.5 | 0.5 | 0.3 | 150 | 7.7 | 22 |
| 10$^4$ | 10$^2$ | 3.5 | 1.5 | 0.3 | 150 | 77 | 2.2 × 10$^2$ |

Table I shows the estimates for the radii of the shell with an extremely small thickness equal to the thick-shell Debye length $L_{D0}$ and the parameters of the optimal laser pulse of the first harmonic of the Nd-laser radiation, which provide acceleration of protons into the target with different characteristic energies $\varepsilon_i$ at a conversion efficiency $\eta = 0.2$. The use of a thin shell that provides amplification of a separated charge field due to multiple oscillations of laser-accelerated electrons can provide electrodynamic implosion of fast ions with energies of several tens of MeV at a relatively small energy of petawatt femtosecond laser pulses of about 10 J. To accelerate ions up to energies of 20–50 MeV, which requires laser pulses with intensities of $10^{21}$–$10^{22}$ W cm$^{-2}$, the pulse duration should be 6–10 fs, and the shell should have a radius of 2–5 $\mu$m and a wall thickness of about 0.3 $\mu$m. To accelerate ions up to energies of 50–150 MeV, laser pulses with intensities of $10^{22}$–$10^{23}$ W cm$^{-2}$ are required, with durations of 3–5 fs. To match the parameters of such an intense pulse and the shell, which provide effective electrodynamic implosion of ions, the pulse energy must already be several tens of joules. Note that the last two rows of Table I present a qualitative estimate; for this and higher intensities, more accurate estimations, accounting e.g. radiation friction, may be required.

## IV. NUMERICAL MODELING AND DISCUSSION OF FAST-ION ELECTRODYNAMIC IMPLOSION

To visualize and support the obtained analytical estimates, PIC simulations were performed with the open-source code SMILEI.[20] For demonstration of the proposed implosion mechanism, a case corresponding to a shell with $R = 7.7$ $\mu$m and $\Delta = 0.3$ $\mu$m, irradiated by a laser pulse with $\tau_L = 14.3$ fs, $E_L = 10$ J, and $I_L = 10^{20}$ W cm$^{-2}$ was selected; see the bold row in Table I. The simulations were performed in a 2D geometry.

The shell was modeled as polyethylene of mass density 0.96 g cm$^{-3}$ with one carbon and two hydrogen atoms. The resolution was 0.01 $\mu$m, and there were 16 electrons, 8 protons, and 4 carbon ions per cell. The interaction scheme is shown in Fig. 2(a), where the absolute value of the electric field is shown along with the propagation direction, indicated by four red arrows. In the center, the dashed circle corresponds to the shell position.

The initial electron density profile is shown in Fig. 2(b). After irradiation, electrons are heated to energies of the order of tens of MeV; see Fig. 2(d) for time 0.13 ps. The hot electron cloud starts to expand and to tire the ions towards the center of the target. This process requires energy, which is transferred from the electrons to become ion kinetic energy. It can be seen from Fig. 2(e) that the electron temperature drops approximately two times during this expansion from 0.13 to 0.2 ps. At this time, as Fig. 2(f) shows, ions are already accelerated to energies of 10–20 MeV at the front of the imploding ion cloud.

The next time instant shown is at 0.5 ps. As can be seen in Figs. 2(g) and 2(h), electrons and ions form a cloud in the central area of the shell. To characterize the level of ion heating near the center of imploding region, an effective temperature is introduced. It is worth noting that the distribution of particles is far from equilibrium, and so it is not correct to define "temperature", but instead the relative stochastic motion of the particles can be described by their average energy in a center-of-mass reference system in a given spatial region. Here, a spatial region for calculating these averages is $\Delta x = \Delta y = 3R/128$. The effective temperature of ions at time 0.5 ps is shown in Fig. 2(i). Both inside and outside the shell, the ions possess a very high effective temperature of the order of MeV. However, according to the density map, this high value corresponds to just a very small fraction of ions at this time. Figure 2(c) shows the characteristic values of the electric field at 0.5 ps, which reach about $10^{12}$ V m$^{-1}$.

The later evolution of the imploding system is shown in Fig. 3, from which it can be seen that the fourfold irradiation works quite symmetrically, with the imprint seen only on the effective temperature map. The density profiles of both ions and electrons look ring-like, and the implosion proceeds rather effectively. According to the results, presented in Figs. 3(a)–3(c), the implosion is not yet finished at 1 ps, and the dense shell is clearly visible. However, the fastest ions have already started to collide in the center, and the effective temperature is about several hundreds of keV. The culmination of the implosion happens at ≈2 ps; see Figs. 3(d)–3(f). At this stage, the densities of the particles reach about half of the initial solid density. The relaxation stage proceeds further, as shown in Figs. 3(g)–3(l), with the density dropping about an order of magnitude on the time scale of 4 ps, and the effective temperature in the center decreases compared with the surrounding area. However, even at 5.8 ps, the ion effective temperature remains a few tens of keV over the whole area of the imploded shell.

To investigate the dependence of the implosion dynamics on the laser and target parameters, additional simulations were performed, with the results shown in Fig. 4. The numerical setup, including the temporal and spatial steps, the number of particles, and the plasma parameters were the same as before. Three sets of initial parameters were used, to demonstrate the analytical scaling (14). As the ions are accelerated toward the center, their energy reaches several tens of MeV just before the collision starts; see Figs. 4(a)–4(c). At this stage, the ions are still cold, and so the effective temperature is very low (not shown). When the collision starts, the collective ion flow at the center transforms to small-scale relative motion of individual particles, resulting in a strong increase in the effective temperature; see Figs. 4(d)–4(f). The scale of this effective temperature is very similar to the kinetic energy of the regular imploding flow before the collision. The three selected cases in Table I correspond to $I_L = 10^{19}$ W cm$^{-2}$ [the first italicized row, Figs. 4(a) and 4(d)], $I_L = 10^{20}$ W cm$^{-2}$ [the bold row, Figs. 4(b) and 4(e)], and $I_L = 10^{21}$ W cm$^{-2}$ [the second italicized row, Figs. 4(c) and 4(f)]. The maximum observed energy and the effective temperatures for these three cases in the simulation results differ by a factor of about two, which is quite close to the analytical scaling (14).

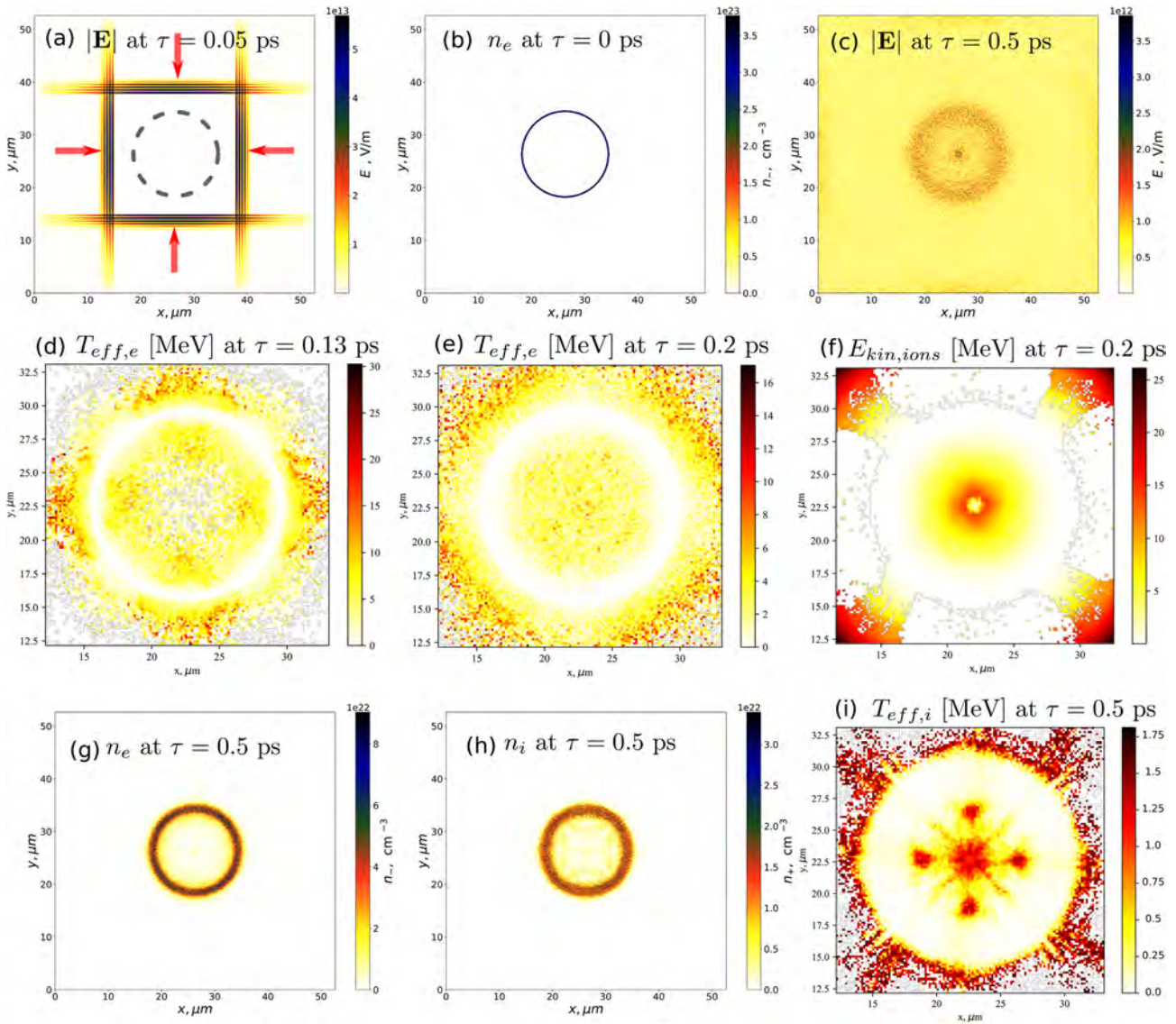

**FIG. 2.** Irradiation scheme and early-time results of 2D PIC simulations for the implosion of a thin plastic shell according to the bold row in Table I. (a) Irradiation scheme showing the laser pulses by the absolute value of the electric field. (b) Initial electron density. (c) Absolute value of the electric field at 0.5 ps. (d) and (e) Electron effective temperature at 0.13 and 0.2 ps, respectively. (f) Ion kinetic energy at 0.2 ps. (g) and (h) Electron and ion densities, respectively, at 0.5 ps. (i) Ion effective temperature at 0.5 ps. Note that the square-fold symmetry evident in (d), (f), and (i) is the result of the fourfold radiation geometry, shown in (a).

The proposed general approach for laser-driven shell implosion can be realized in the cylindrical or spherical irradiation geometry.

Note that although the 2D (cylindrical) case considered here looks somewhat less promising in terms of high compression and temperature levels than the 3D (spherical) one, it may be difficult to achieve implosions with significantly greater symmetry in the 3D situation using standard laser-driven irradiation schemes. Besides, in practice, the open cylindrical scheme is more relevant for diagnostics.

It is important to note that because of the kinetic nature of the electrodynamic implosion, no hydrodynamic instabilities develop. However, there is a need for more detailed studies examining those features that are not directly predicted by the analytical consideration, such as degree of compression, dependence on laser focusing, and sensitivity to nonidealities of the layers and irradiation.

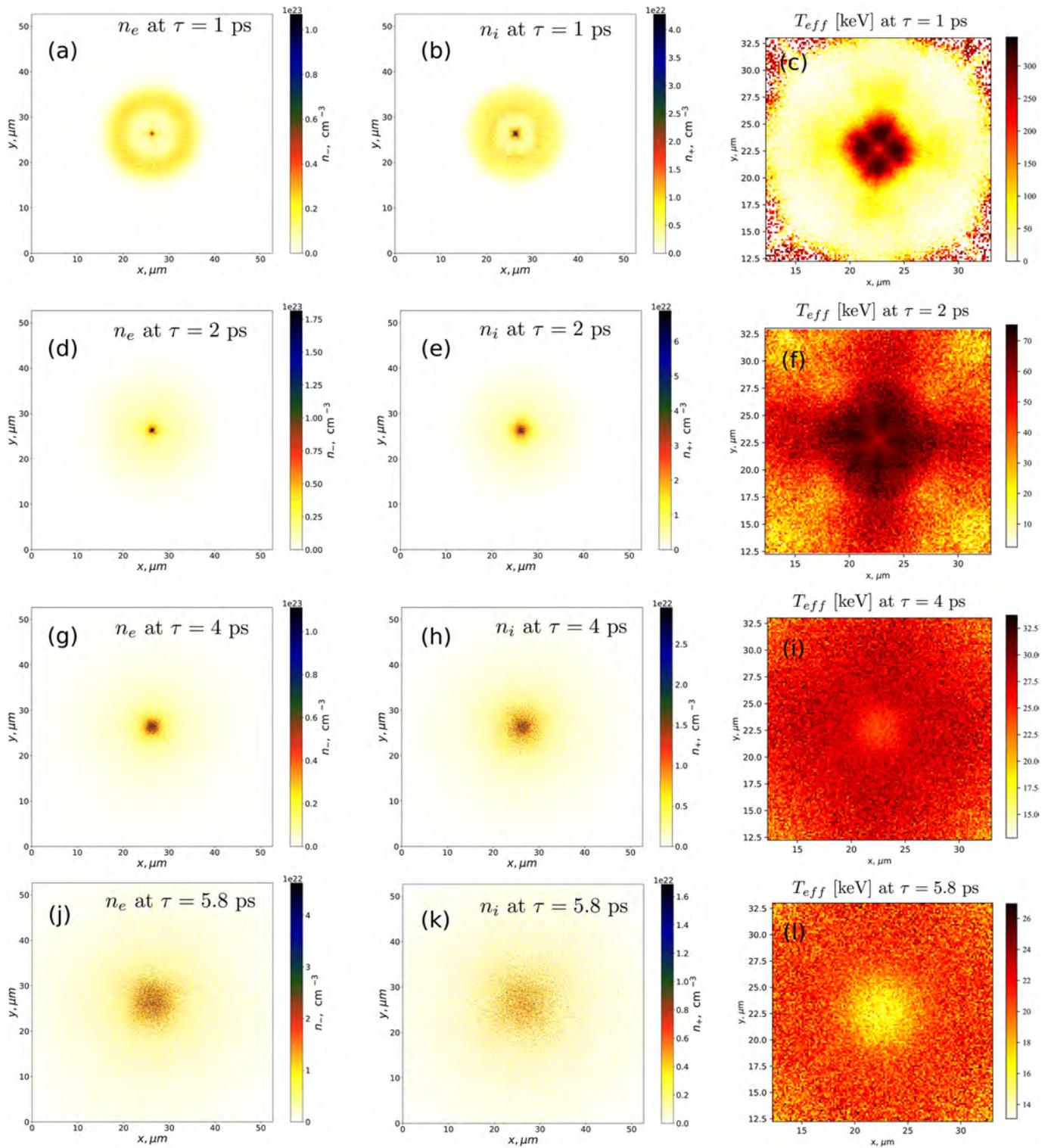

**FIG. 3.** Results of 2D PIC simulations of the implosion following irradiation of a thin plastic shell according to the bold row in Table I. (a), (d), (g), and (j) Electron density at times 1, 2, 4, and 5.8 ps, respectively. (b), (e), (h), and (k) Ion density at times 1, 2, 4, and 5.8 ps, respectively. (c), (f), (j), and (l) Effective ion temperature at times 1, 2, 4, and 5.8 ps respectively. The fourfold symmetry evident in (c), (f), (i), and (l) is a consequence of the fourfold irradiation scheme, shown in Fig. 2(a).

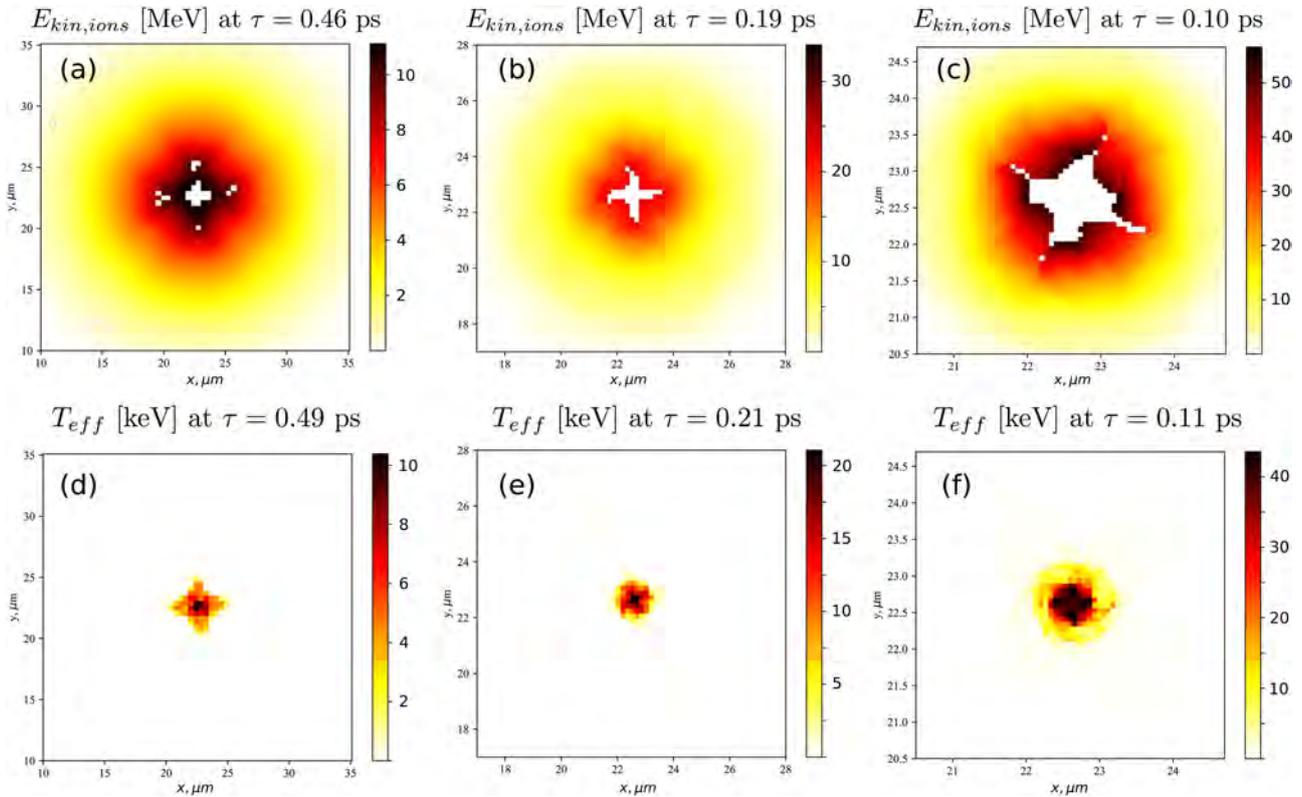

**FIG. 4.** Proton energies just before collision [(a)–(c)] and effective temperatures just after collision [(d)–(f)] for the three cases in Table I: the first italicized row, with $I_L = 10^{19}$ W cm$^{-2}$ [(a) and (d)]; the bold row, with $I_L = 10^{20}$ W cm$^{-2}$ [(b) and (e)]; and the second italicized row, with $I_L = 10^{21}$ W cm$^{-2}$ [(c) and (f)]. After the collision, the energy is transformed to the effective temperature with approximately the same value.

## V. CONCLUSIONS

The electrodynamic implosion of fast ions inside a hollow thin-walled shell irradiated with short petawatt laser pulses has been theoretically investigated. To ensure the maximum implosion efficiency, it is proposed to use a shell with wall thickness close to the Debye length of laser-accelerated fast electrons. In this case, the numerous oscillations of fast electrons near the shell wall lead to a significant increase in the magnitude of the separated charge field. It is found that irradiation of such a shell with a radius of about 10 $\mu$m and a wall thickness of 0.2–0.4 $\mu$m by laser pulses with an intensity of $10^{21}$–$10^{23}$ W cm$^{-2}$ and a duration of several femtoseconds may provide implosion of ions with energies up to 100 MeV, leading to an effective ion temperature in the MeV range after collision, with a relatively small laser pulse energy ~10 to 100 J. In this case, the flux density of the imploded ions may reach $10^{30}$–$10^{31}$ particles cm$^{-2}$ s$^{-1}$. Electrodynamic implosion and collision of such dense flows of ions with energies of several tens of MeV and more represents an extremely interesting topic for experimental research in high-energy-density physics.

## ACKNOWLEDGMENTS

This work was financially supported by the Russian Science Foundation under Project No. 21-11-00102. The authors acknowledge the NRNU MEPhI High-Performance Computing Center and the Joint Supercomputer Center RAS for conducting the numerical simulations. The authors are grateful to I. Ya. Doskoch for help in preparing the manuscript.

## AUTHOR DECLARATIONS

### Conflict of Interest

The authors have no conflicts to disclose.

### Author Contributions

**S. Yu. Gus'kov**: Conceptualization (equal); Data curation (equal); Formal analysis (equal); Funding acquisition (equal); Investigation (equal); Methodology (equal); Project administration (equal); Supervision (equal); Validation (equal); Writing – original draft (equal); Writing – review & editing (equal). **Ph. Korneev**: Data curation (equal); Formal analysis (equal); Investigation (equal); Methodology (equal); Software (equal); Visualization (equal); Writing – original draft (equal); Writing – review & editing (equal). **M. Murakami**: Data curation (supporting); Formal analysis (supporting); Investigation (equal); Methodology (equal); Validation (equal); Writing – review & editing (equal).

## DATA AVAILABILITY

The data that support the findings of this study are available on a reasonable request from the corresponding author.